\documentclass[floatfix,%
 reprint,
superscriptaddress,
twocolumn,
nofootinbib,
 amsmath,amssymb,
 aps,
]{revtex4-2}

\usepackage{graphicx}
\usepackage{dcolumn}
\usepackage{bm}
\usepackage{breqn}
\usepackage{xcolor}
\usepackage{ragged2e}
\usepackage{amsmath}
\usepackage[title]{appendix}
\usepackage[normalem]{ulem}

\begin{document}

\title{Quantum oscillations and anisotropic magnetoresistance in the quasi-two-dimensional Dirac nodal line superconductor $\mathrm{YbSb_2}$}

\author{Yuxiang Gao}
\affiliation{Department of Physics and Astronomy$,$ Rice University$,$ Houston$,$ Texas 77005$,$ USA}
\affiliation{Rice Center for Quantum Materials$,$ Rice University$,$ Houston$,$ Texas 77005$,$ USA}
\author{Kevin Allen}
\affiliation{Department of Physics and Astronomy$,$ Rice University$,$ Houston$,$ Texas 77005$,$ USA}
\affiliation{Rice Center for Quantum Materials$,$ Rice University$,$ Houston$,$ Texas 77005$,$ USA}
\author{Rose Albu Mustaf}
\affiliation{Department of Physics and Astronomy$,$ Rice University$,$ Houston$,$ Texas 77005$,$ USA}
\affiliation{Rice Center for Quantum Materials$,$ Rice University$,$ Houston$,$ Texas 77005$,$ USA}
\author{Yichen Zhang}
\affiliation{Department of Physics and Astronomy$,$ Rice University$,$ Houston$,$ Texas 77005$,$ USA}
\affiliation{Rice Center for Quantum Materials$,$ Rice University$,$ Houston$,$ Texas 77005$,$ USA}
\author{Sanu Mishra}
\affiliation{Department of Physics and Astronomy$,$ Rice University$,$ Houston$,$ Texas 77005$,$ USA}
\affiliation{Rice Center for Quantum Materials$,$ Rice University$,$ Houston$,$ Texas 77005$,$ USA}
\author{Christopher Lane}
\affiliation{Theoretical Division$,$ Los Alamos National Laboratory$,$ Los Alamos$,$ New Mexico 87545$,$ USA }
\author{Marta Zonno}
\affiliation{Canadian Light Source$,$ Inc.$,$ 44 Innovation Boulevard$,$ Saskatoon$,$ SK$,$ S7N 2V3$,$ Canada}
\author{Sergey Gorovikov}
\affiliation{Canadian Light Source$,$ Inc.$,$ 44 Innovation Boulevard$,$ Saskatoon$,$ SK$,$ S7N 2V3$,$ Canada}
\author{Jian-Xin Zhu}
\affiliation{Theoretical Division$,$ Los Alamos National Laboratory$,$ Los Alamos$,$ New Mexico 87545$,$ USA }
\affiliation{Center for Integrated Nanotechnologies$,$ Los Alamos National Laboratory$,$ Los Alamos$,$ New Mexico 87545$,$ USA}
\author{Ming Yi}
\affiliation{Department of Physics and Astronomy$,$ Rice University$,$ Houston$,$ Texas 77005$,$ USA}
\affiliation{Rice Center for Quantum Materials$,$ Rice University$,$ Houston$,$ Texas 77005$,$ USA}
\author{Emilia Morosan}
\email[corresponding author: E. Morosan ]{emorosan@rice.edu}
\affiliation{Department of Physics and Astronomy$,$ Rice University$,$ Houston$,$ Texas 77005$,$ USA}
\affiliation{Rice Center for Quantum Materials$,$ Rice University$,$ Houston$,$ Texas 77005$,$ USA}

\date{\today}

\begin{abstract}

Recent interest in quantum materials has focused on systems exhibiting both superconductivity and non-trivial band topology as material candidates to realize topological or unconventional superconducting states. So far, superconductivity in most topological materials has been identified as type II. In this work, we present magnetotransport studies on the quasi-two-dimensional type I superconductor $\mathrm{YbSb_2}$. Combined ab initio DFT calculations and quantum oscillation measurements confirm that $\mathrm{YbSb_2}$ is a Dirac nodal line semimetal in the normal state. The complex Fermi surface morphology is evidenced by the non-monotonic angular dependence of both the quantum oscillation amplitude and the magnetoresistance. Our results establish $\mathrm{YbSb_2}$ as a candidate material platform for exploring the interplay between band topology and superconductivity.

\end{abstract}

\maketitle

\section{Introduction}


The coexistence of non-trivial electronic band topology and unconventional superconductivity represents a cutting-edge area of condensed matter research, yet its experimental realization remains challenging and rare. Superconductivity and non-trivial topology have been reported together in a few material systems, such as kagome compounds $\mathrm{(K,Rb,Cs)V_3Sb_5}$ \cite{Ortiz2019, Ortiz2020, Hu2022}, pyrochlore $\mathrm{CeRu_2}$ \cite{Matthias1958,Kiss2005,Huang2024}, twisted graphene moiré and rhombohedral stacked graphene systems \cite{Bistritzer2011,Cao2018,Henck2018,Oh2021,Lisi2021,Zhou2021,ZhangH2024,Han2025}, intercalated transition metal dichalcogenides PbTaSe$_2$, In$_x$TaS$_2$, In$_x$TaSe$_2$ \cite{Ali2014,Bian2016,Li2021,Zhang2024}, all type II superconductors. Several other well-known type II superconductors have also recently been revisited for their band topology, including $\mathrm{MgB_2}$ \cite{Nagamatsu2001,Hinks2001,Jin2019,Zhou2019} and the iron-based superconductors \cite{Kamihara2006,Kamihara2008,Wen2008,Hsu2008,Pitcher2008,Wu2016,Zhang2018,Zhang2019}. $\mathrm{PdTe_2}$ had been reported to be a type I superconductor with type II Dirac points \cite{Leng2019,Noh2017,Das2018,Clark2018}, but recent $\mu$SR studies suggest that its superconductivity is type II \cite{Gupta2024}. In spite of these efforts, the role of non-trivial band topology in type II superconductors remains elusive. Moreover, type I superconductors might provide a distinct platform for the exploration of the interplay between topology and superconductivity. Given the absence(presence) of a vortex state in a type I(II) superconductor, the role of non-trivial band topology is likely different in the two classes of compounds, motivating the interest in type I superconductivity in topological materials. However, no compounds exhibiting \textit{both} non-trivial band topology and type I superconductivity have been identified, primarily because type I superconductors are typically elemental, while non-trivial band topology is mostly realized in compounds.

Here, we identify $\mathrm{YbSb_2}$ as the first compound exhibiting both type I superconductivity \textit{and} non-trivial band topology. $\mathrm{YbSb_2}$ is a known type I superconductor \cite{Zhao2012}, but its topology has not yet been investigated. We used anisotropic magnetoresistance (AMR) and quantum oscillation (QO) measurements, together with band structure calculations, to demonstrate its non-trivial topology, with a Dirac nodal line and its magnetic breakdown orbit identified in the fast Fourier transform (FFT) spectra. 


Previous thermodynamic and transport measurements have revealed a superconducting state with T$_c$ = 1.5 K in $\mathrm{YbSb_2}$, with radio frequency susceptibility measurements hinting at a possible second superconducting state below T$_c^{(2)}$ = 0.4 K \cite{Zhao2012}. In contrast to the superconducting state, the normal state of $\mathrm{YbSb_2}$ has barely been examined, with the exception of a de Haas-van Alphen study \cite{Sato1999}. Here we focus on the normal state and its Fermi surface topology as determined from QO measurements and density functional theory (DFT) calculations. Both experiments and calculations reveal a Dirac nodal line in the electronic band structure. QOs arising from the Dirac nodal line and its magnetic breakdown orbit are identified in the FFT spectra. Despite the quasi-two-dimensional crystal structure of $\mathrm{YbSb_2}$, the angle-dependent QO (AQO) amplitude and anisotropic magnetoresistance (AMR) show non-monotonic angular dependence, which is attributed to the switching between different underlying Fermi surface cross-sections. In the end, we proposed a possible connection between the Dirac nodal lines and the type I superconductivity in YbSb$_2$.
%



\begin{figure*}
\includegraphics[width=0.95\textwidth]{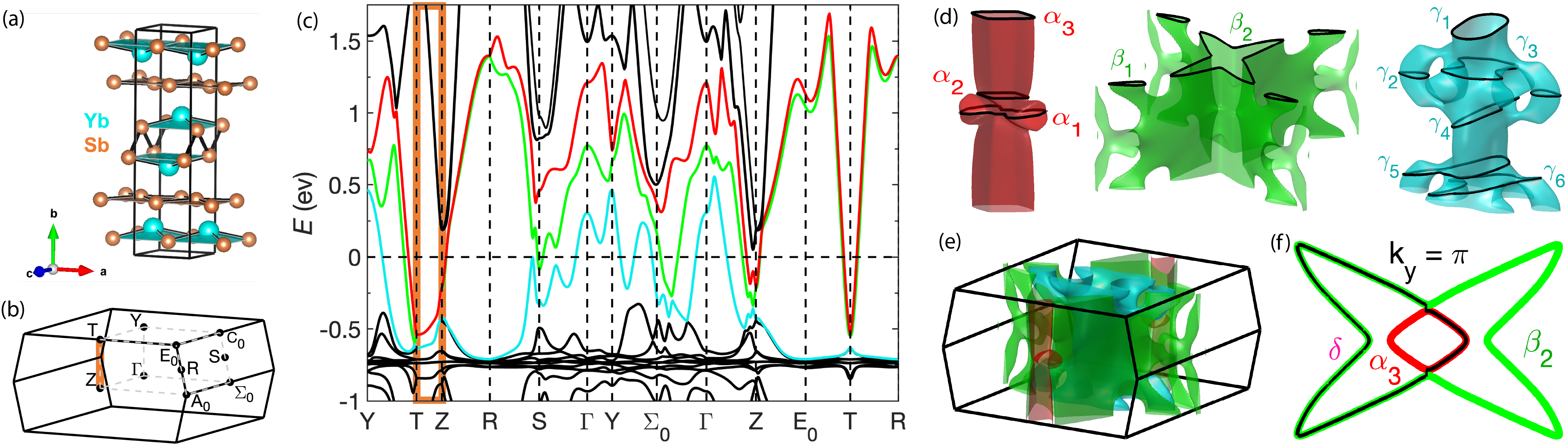} 
\caption{\label{fig:1} Crystal structure, band structure, and Fermi surface of $\mathrm{YbSb_2}$. (a) Crystal structure of $\mathrm{YbSb_2}$. (b) First Brillouin zone of $\mathrm{YbSb_2}$. The direction of the Dirac nodal line is highlighted in orange. (c) Electronic band structure of $\mathrm{YbSb_2}$ from DFT calculations. The bands that cross the Fermi energy are highlighted in red, green, and cyan. (d) Calculated Fermi pockets of $\mathrm{YbSb_2}$ with SOC included. The colors of the Fermi pockets match those in (c), and the labels represent the QO frequencies as shown in Fig.~\ref{fig:2}-\ref{fig:5}. (e) Calculated Fermi surface of $\mathrm{YbSb_2}$ in the first Brillouin zone, illustrating the positions of the Fermi pockets in (c). (f) Magnetic breakdown orbit $\delta$ formed by $\alpha_3$ and $\beta_2$ cross-sections. }
\end{figure*} 

\section{Methods}
$\mathrm{YbSb_2}$ single crystals were synthesized using a flux growth technique as described in \cite{Zhao2012}. 
The electrical transport measurements were conducted in a Quantum Design (QD) Dynacool PPMS system with a 14 T magnet. The resistance was measured using the electrical transport option (ETO). Magnetization up to 9 T was measured in the QD Dynacool PPMS using the vibrating sample magnetometer (VSM) option.

The $\mathrm{YbSb_2}$ electronic band structure was obtained based on DFT using the Vienna ab-initio simulation package (VASP) \cite{vasp} using the Perdew-Berke-Ernzerhof (PBE) exchange-correlation with the generalized-gradient approximation (GGA) \cite{Kresse1999,Perdew1996}. To account for the localized \textit{f} electrons, an on-site Hubbard U = 4 eV was applied to the Yb \textit{4f} states. A Wannier tight-binding model was fitted to the electronic band structure via the Wannier90 package \cite{Pizzi2020}.

Angle-resolved photoemission spectroscopy (ARPES) measurements were performed at the QMSC beamline of the Canadian Light Source equipped with a R4000 electron analyzer and using $p$-polarized light. The YbSb$_2$ samples were cleaved in-situ and measured at T = 15 K, under vacuum conditions better than 6$\times$10$^{-11}$ torr. 

\section{Results}

As shown in Fig. \ref{fig:1}(a), the $\mathrm{YbSb_2}$ unit cell can be viewed as a stack of [Yb-Sb] - [Sb] - [Yb-Sb] layers along the {\it b} axis, resulting in a quasi-two-dimensional orthorhombic crystal structure. The electronic band structure determined from first-principles DFT+{\it U} calculations along with the first Brillouin zone are shown in Figs. \ref{fig:1}(b,c). To estimate the U for YbSb$_2$, we carried out ARPES measurements (Fig. \ref{fig:SI1}), from which we determined that the energy of the \textit{4f} electronic bands was at $\mathrm{E-E_F}\sim-$ 0.6 eV. To match the energy of these \textit{4f} electronic bands, an on-site Hubbard {\it U} = 4 eV is applied.  In $\mathrm{YbSb_2}$, three bands [red, green, and cyan in panel (c)] cross the Fermi energy and give way to a complex set of Fermi surfaces, as shown in Figs. \ref{fig:1}(d,e). Each Fermi pocket in Fig. \ref{fig:1}(d) exhibits quasi-two-dimensional features (elongation along the $k_y$ axis) and three-dimensional features (heavy corrugation). Due to their complex shape, each Fermi pocket features multiple extremal cross sections [labeled in Fig. \ref{fig:1}(d)] that may contribute to the QOs. Additionally, as shown in Fig. \ref{fig:1}(e), the orbits $\alpha_3$ (red) and $\beta_2$ (green) are close to each other, potentially giving rise to a magnetic breakdown orbit $\delta$ under a strong magnetic field \cite{Shoenberg1984}. A schematic of the orbits $\alpha_3$, $\beta_2$, and $\delta$ is shown in Fig. \ref{fig:1}(f). Moreover, since both $\alpha$ and $\beta$ are electronic bands, the cross-sectional areas $S_i$ (i = $\delta$, $\alpha_3$, $\beta_2$) are related through the relation: 
\begin{equation*}
    S_\delta=(S_{\alpha_3}+S_{\beta_2})/2.
\end{equation*}

Highlighted in orange in Fig. \ref{fig:1}(b,c), the bands along the T-Z direction form Dirac nodal lines. Therefore, the Fermi pockets $\alpha_1$, $\alpha_2$, $\alpha_3$, and the magnetic breakdown orbit $\delta$ are directly related to the Dirac nodal lines. and the underlying quasiparticles from these Fermi pockets may influence magnetotransport properties, such as magnetoresistance and QO.

\begin{figure} [b]
\includegraphics[width=0.48\textwidth]{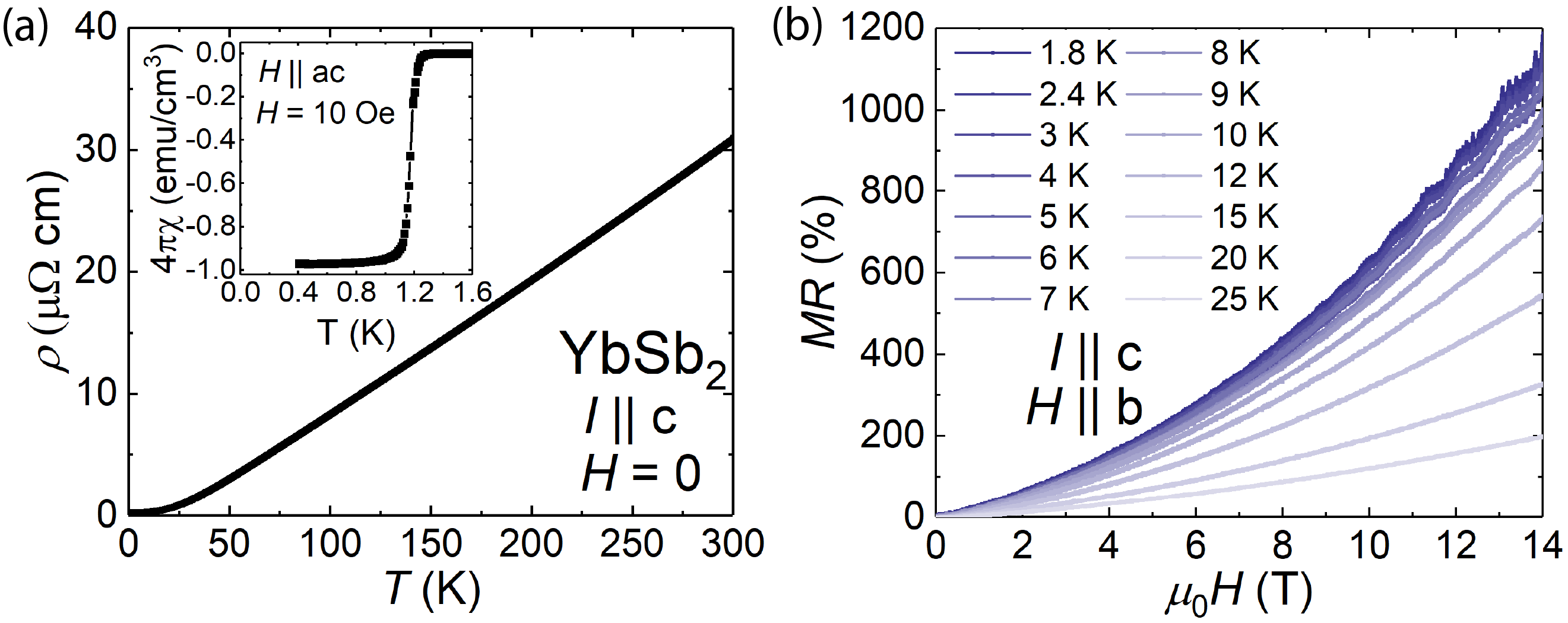} 
\caption{\label{fig:1_2} Resistivity, susceptibility, and magnetoresistance (MR) of $\mathrm{YbSb_2}$. (a) Zero field resistivity of $\mathrm{YbSb_2}$ as a function of temperature. The inset is the susceptibility of $\mathrm{YbSb_2}$ illustrating the bulk superconductivity. (b) Magnetoresistance (MR) of $\mathrm{YbSb_2}$ at different temperatures for current along c-axis and magnetic field along b-axis.}
\end{figure}

\begin{figure*} 
\includegraphics[width=0.96\textwidth]{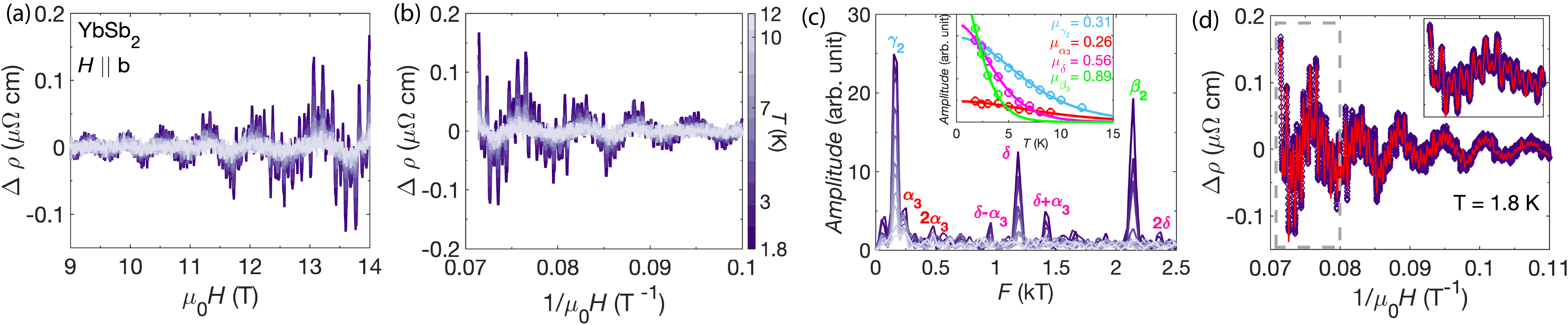} 
\caption{\label{fig:2}Shubnikov-de Haas (SdH) oscillations of $\mathrm{YbSb_2}$ with $\mathbf{H} \parallel \mathbf{b}$. (a,b) SdH oscillations of $\mathrm{YbSb_2}$ as a function of magnetic field $\mu_0 H$ (a) and inverse magnetic field $1/\mu_0 H$ (b) at different temperatures. (c) Fast Fourier transform (FFT) spectra of (b). The inset shows the fit to the Lifshitz-Kosevich (LK) thermal damping term to extract the effective masses of different frequencies. (d) LK fitting (red line) to the SdH oscillations (purple symbols) at 1.8 K. The inset shows the high-frequency SdH oscillations and fitting in the region enclosed by the grey box. } 
\end{figure*}
\begin{figure*} 
\includegraphics[width=0.96\textwidth]{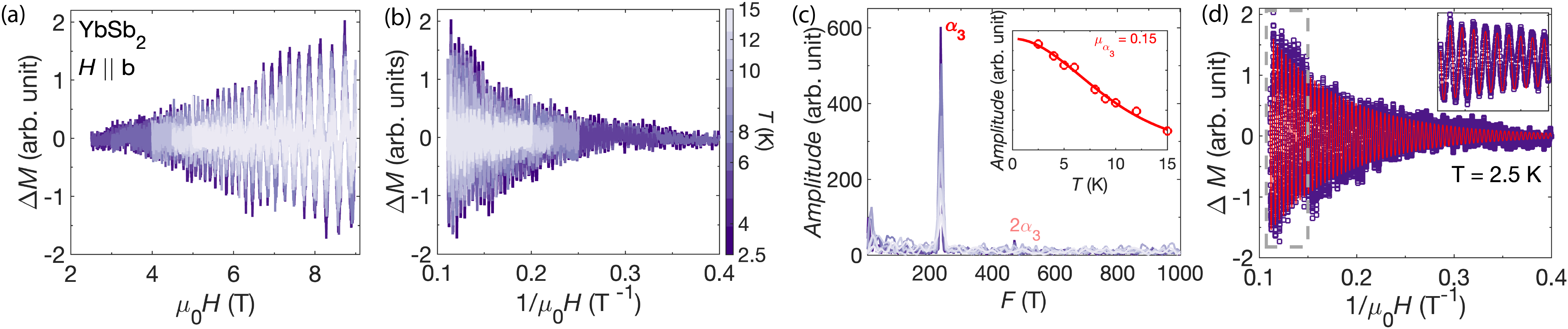} 
\caption{\label{fig:3}de Haas-van Alphen (dHvA) oscillations of $\mathrm{YbSb_2}$ with $\mathbf{H} \parallel \mathbf{b}$. (a,b) dHvA oscillations of $\mathrm{YbSb_2}$ as a function of magnetic field $\mu_0 H$ (a) and inverse magnetic field $1/\mu_0 H$ (b) at different temperatures. (c) FFT spectra of (b). The inset shows the fit to the LK thermal damping term to extract the effective masses corresponding to the $\alpha_3$ frequencies. (d) LK fitting (red line) to the dHvA oscillations (purple symbols) at 2.5 K. The inset shows the dHvA oscillations and fitting in the region enclosed by the grey box.} 
\end{figure*}

The $\mathrm{YbSb_2}$ zero-field resistivity as a function of temperature is shown in Fig. \ref{fig:1_2}(a). The large residual resistivity ratio RRR = $\rho(300\,$K$)/\rho(1.8\,$K$)$ = 110 indicates high crystal quality. The susceptibility [inset, Fig. \ref{fig:1_2}(a)] confirms a bulk superconducting state below $T_c$ = 1.3 K for {\it H} = 10 Oe, consistent with previous studies \cite{Zhao2012, Sato1999}. The field-dependent resistivity measurements reveal strong Shubnikov de-Haas (SdH) oscillations and large, non-saturating MR $\simeq$ 1200\% at 1.8 K and 14 T [Fig. \ref{fig:1_2}(b)]. This is commonly seen as an indicator of non-trivial band topology \cite{Liang2015,Shekar2015,Hu2019}. 

The SdH oscillations are more visible in $\Delta \rho(H)$, which is the magnetoresistance after a smooth background is subtracted. The $\Delta \rho(H)$ manifold for different temperatures is shown in Figs. \ref{fig:2}(a,b). The FFT spectra of the SdH oscillations reveal several frequencies [Fig. \ref{fig:2}(c)], consistent with the complex Fermi surface shown in Figs. \ref{fig:1}(d,e). The SdH oscillations are further fit with the Lifshitz-Kosevich (LK) formula \cite{Shoenberg1984}:
\begin{multline} 
\label{eq1} 
\frac{\Delta \rho}{\rho_0} \propto B^\gamma \sum_p \sum_{r=1}^{\infty}\frac{1}{r^{1/2}}R_T R_D R_S \\ 
\cos\Bigg(2\pi \bigg[r\bigg(\frac{F}{B}-c+b\bigg)+d\bigg]\Bigg), 
\end{multline}
where $R_T={raT \mu}/$sinh$(raT \mu)$, $R_D=\exp(-{raT_D\mu}/{B})$, and $R_S=\cos({r\pi g\mu}/{2})$ are the temperature, field, and spin damping factors, respectively. $p$ and $r$ are the QO frequency and harmonic indices, $\mu={m_{\texttt{eff}}}/{m_0}$ is the ratio of effective mass $m_{\texttt{eff}}$ to free electron mass $m_0$, $a=(2{\pi}^2 k_B m_0)/ (\hbar e)\approx14.69$ T/K, $T_D$ is the Dingle temperature and $F$ is the frequency. $c=0,~d=0$ for a 2D pocket and $c=1/2,~d=\pm 1/8$ for a minimum/maximum cross section of a 3D electron/hole pocket, respectively. $b~=~\phi_B/2\pi$ where $\phi_B$ is the Berry phase. An example of the fit to the SdH oscillations at 1.8 K is shown in Fig. \ref{fig:2}(d) (red line), and by fitting the QO amplitude of each frequency at different temperatures to the temperature-damping term $R_T$, a small effective mass is obtained [see the inset of Fig. \ref{fig:2}(c)]. The parameters extracted from the LK formula are listed in Table \ref{table:1}. 

\begin{figure*} 
\includegraphics[width=0.96\textwidth]{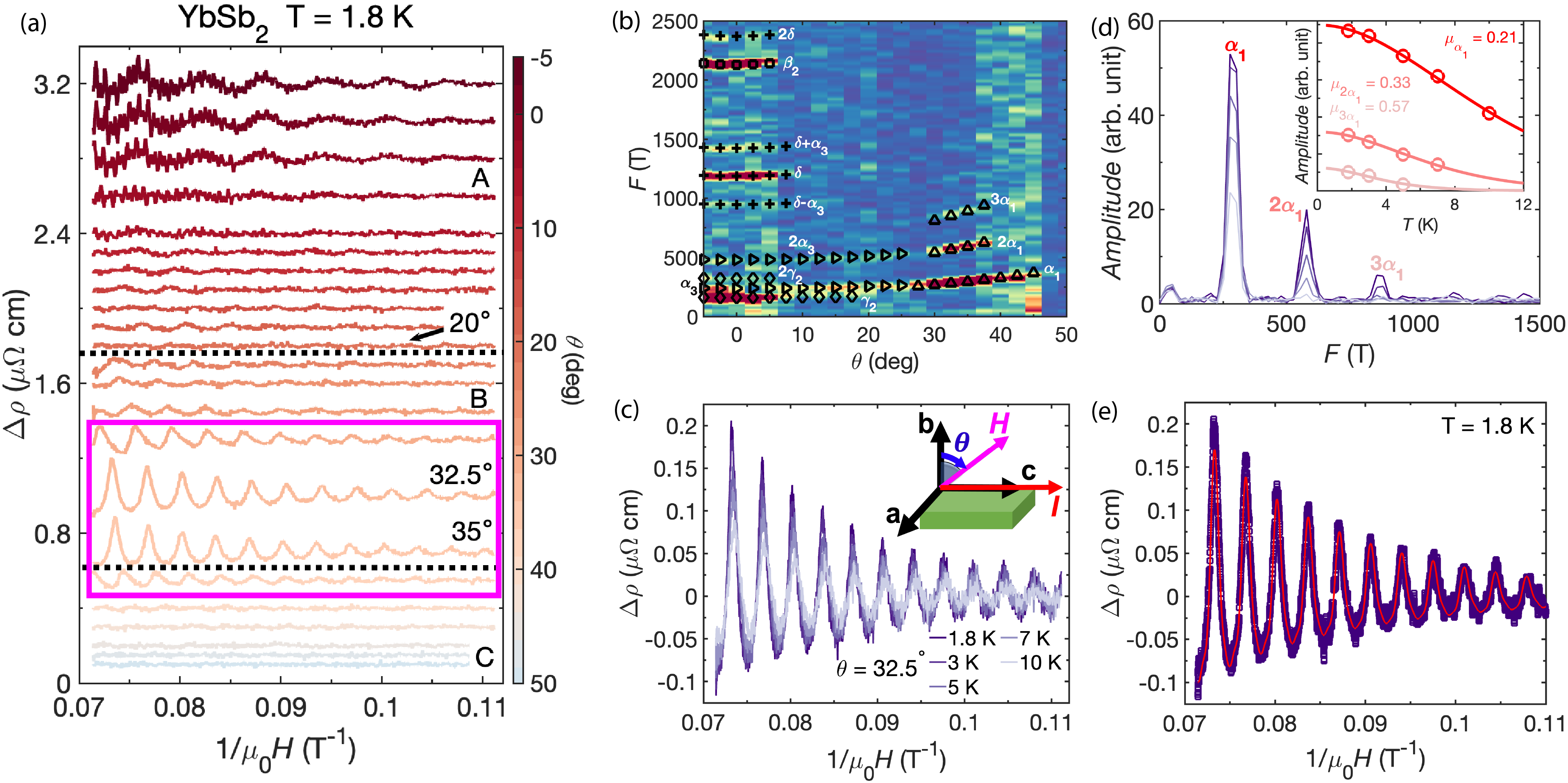} 
\caption{\label{fig:4} 
Angle-dependent quantum oscillations (AQO) of $\mathrm{YbSb_2}$ with the magnetic field in the bc-plane. (a) Waterfall plot of the SdH oscillations after background subtraction. A vertical offset is applied for better visualization. The oscillations highlighted in the magenta rectangle show a non-sinusoidal-like waveform. The dotted lines divide the oscillations into three regions depending on the QO amplitude. (b) Contour plot of the FFT spectra of (a) with the calculated cross-sectional area from DFT (symbols). The contour represents the FFT intensities, from low (blue) to high (red).  (c) SdH oscillations at different temperatures for $\theta = 32.5^\circ$.  The inset shows the measurement setup for AQO. (d) FFT spectra of (c). The inset shows the fit to the Lifshitz-Kosevich (LK) thermal damping term used to extract the effective masses of $\alpha_1$ and its higher harmonics. (e) LK fitting (red line) to the SdH oscillations (purple symbols) at 1.8 K. }
\end{figure*}

In addition to the SdH oscillations, the de Haas-van Alphen (dHvA) oscillations are observed in the magnetization of $\mathrm{YbSb_2}$ [Figs. \ref{fig:3}(a,b)].  dHvA oscillations were also reported in \cite{Sato1999}. The FFT spectra obtained from Figs. \ref{fig:3}(a,b) consist of only two frequencies [Fig. \ref{fig:3}(c)], $\alpha_3$ and $2 \alpha_3$, in contrast to the complex FFT spectra with more than two frequencies in \cite{Sato1999}, including the two observed in the present study. The difference in FFT spectra could be attributed to the difference in base temperature between this work (2.5 K) and \cite{Sato1999} (0.5 K), with the amplitudes of some QOs being too small to observe at higher T due to the large effective masses. The dHvA oscillations are further fitted using the LK formula for magnetization \cite{Shoenberg1984}: \begin{multline}
\label{eq2}
\Delta M \propto -B^\gamma \sum_p \sum_{r=1}^{\infty}\frac{1}{r^{3/2}}R_T R_D R_S \\
\sin\Bigg(2\pi \bigg[r\bigg(\frac{F}{B}-c+b\bigg)+d\bigg]\Bigg)
\end{multline}
An example of the fit to the dHvA oscillations at 2.5 K is presented in Fig. \ref{fig:3}(d) (red line), with the effective mass fitting for the pocket $\alpha_3$ revealing a small mass of 0.15 $m_e$, similar to the value of 0.115 $m_e$ reported in \cite{Sato1999}. 

\begin{figure*} 
\includegraphics[width=0.96\textwidth]{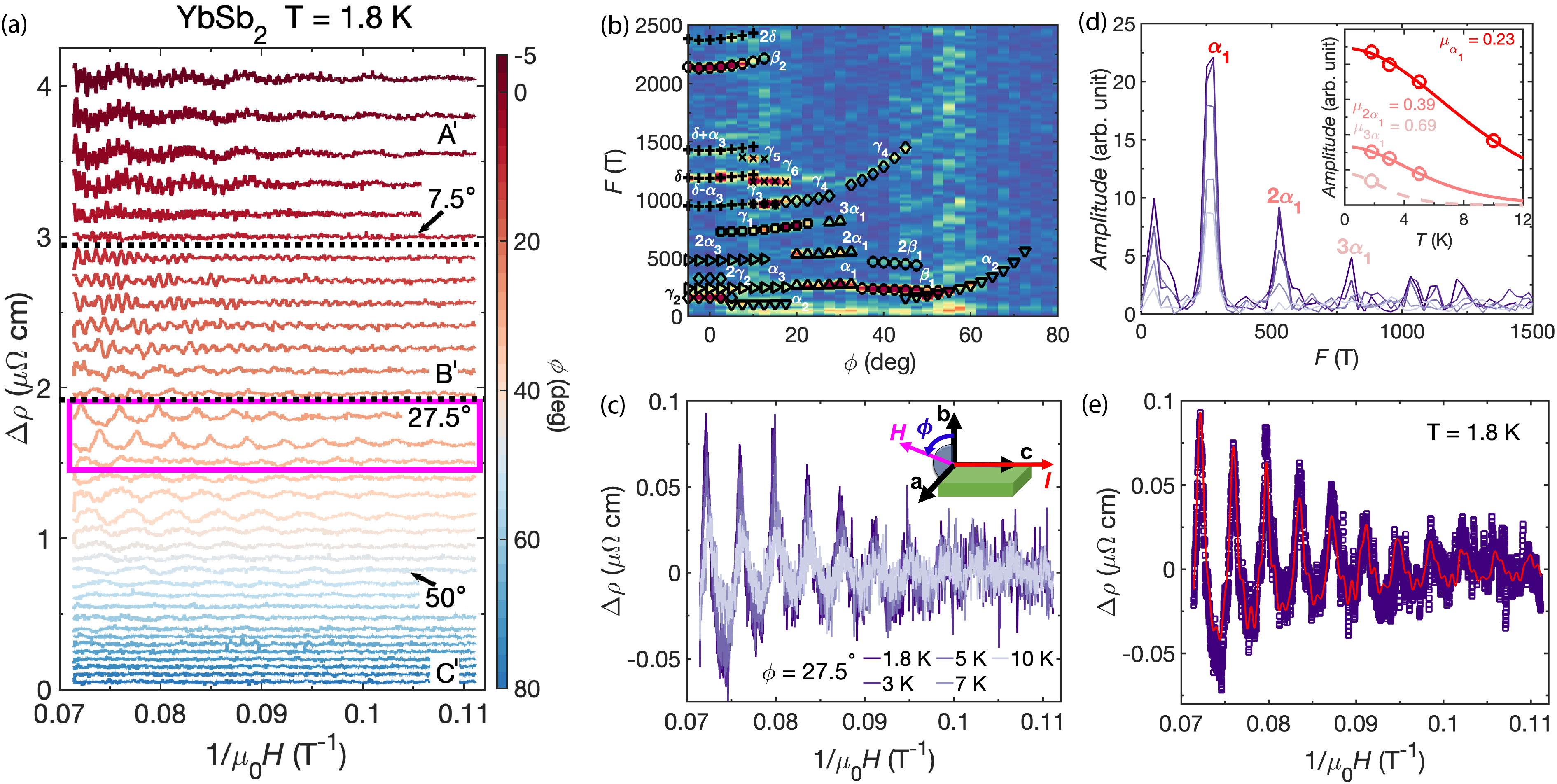} 
\caption{\label{fig:5} 
Angle-dependent quantum oscillations (AQO) of $\mathrm{YbSb_2}$ with the magnetic field in the ab-plane. (a) Waterfall plot of the SdH oscillations after background subtraction. A vertical offset is applied for better visualization. The oscillations highlighted in the magenta rectangle show a non-sinusoidal-like waveform. The dotted lines divide the oscillations into three regions depending on the QO amplitude. (b) Contour plot of the FFT spectra of (a) with the calculated cross-sectional area from DFT (symbols). The contour represents the FFT intensities, from low (blue) to high (red). (c) SdH oscillations at different temperatures for $\phi = 27.5^\circ$. The inset shows the measurement setup for AQO. (d) FFT spectra of (c). The inset shows the fit to the LK thermal damping term used to extract the effective masses of $\alpha_1$ and its higher harmonics. (e) LK fitting (red line) to the SdH oscillations (purple symbols) at 1.8 K.} 
\end{figure*}

The effective mass of the $\alpha_3$ pocket extracted from the SdH oscillations is notably larger than that of the dHvA oscillations. This difference in effective mass has been observed in low-dimensional systems such as layered organic superconductors \cite{Kartsovnik2004} and the Dirac semimetal SrMnSb$_2$ \cite{Liu2017}. The difference can be explained by the underlying mechanisms of the two effects. The dHvA effect originates from oscillations in the free energy of the electrons, where the temperature and field damping strictly involve single-particle processes. In contrast, SdH oscillations arise from oscillations in resistivity, which are caused by variations in the scattering rate. The temperature and field damping in SdH oscillations may be enhanced by additional scattering processes, leading to an apparent increase in the effective mass compared to the dHvA oscillations.

To further investigate the Fermi surface of $\mathrm{YbSb_2}$, we conducted AQO measurements. The QO frequency F is proportional to the extremal Fermi surface cross-section perpendicular to the magnetic field $S_{ext}$, through the Onsager relationship: 
\begin{equation}
  F=\frac{\hbar S_{ext}}{2\pi e}  
\end{equation} where $\hbar$ is the reduced Planck constant. Fig. \ref{fig:4}(a) shows the SdH oscillations as a function of the inverse magnetic field with different field orientations $\theta$ from {\it H} $\parallel$ b ($\theta$ = 0) towards {\it H} $\parallel$ c at 1.8 K, in steps of 2.5$^\circ$. A sketch of the measurement setup is shown in the inset of Fig. \ref{fig:4}(c). 

Strong SdH oscillations are observed up to $\theta = 45^\circ$. However, the amplitude of the SdH oscillations does not change monotonically when the field moves away from H $\parallel b$, as would be expected for a quasi-two-dimensional system. Instead, based on the evolution of the QO amplitude with increasing  $\theta$, the waterfall plot in Fig. \ref{fig:4}(a) can be divided into three regions: the QO amplitude (A) decreases for $\theta \leq 20^\circ$; (B) increases rapidly for $25^\circ \leq \theta \leq 35^\circ$; and (C) decreases for $\theta > 35^\circ$, until it becomes too small to observe when $\theta \geq 45^\circ$. The boundaries for these three regions are indicated by dotted lines in Fig. \ref{fig:4}(a). Two specific $\theta$ values are highlighted for later discussion, where the QO amplitude reaches a local minimum (maximum): $\theta_1 = 20^\circ$ ($\theta_2 = 35^\circ$). The evolution of the QO frequencies $F$ with respect to $\theta$ is plotted in Fig. \ref{fig:4}(b) as a contour plot. 

In addition to the non-monotonic change in the QO amplitude at $30^\circ \leq \theta \leq 37.5^\circ$ [magenta box in Fig. \ref{fig:4}(a)], the waveform of these oscillations appears to deviate from a sinusoidal function, although the periodicity in $1/H$ is preserved. To gain more insight into these oscillations, the SdH oscillations at $\theta = 32.5^\circ$ were measured at different temperatures [Fig. \ref{fig:4}(c)]. The waveform of the SdH oscillations becomes more sinusoidal at higher temperatures (for example, {\it T} = 10 K). The FFT spectra of the SdH oscillations at $\theta = 32.5^\circ$ for different temperatures are shown in Fig. \ref{fig:4}(d). The higher harmonics of the fundamental oscillations with frequency $\alpha_1$ are observed at T $<$ 10 K, while these harmonics are absent at {\it T} = 10 K. The superposition of strong higher harmonics onto fundamental oscillations alters the appearance of the SdH oscillations, making them non-sinusoidal. This is confirmed in Fig. \ref{fig:4}(e), where the SdH oscillation data at 1.8 K (purple symbols) are well-fit with the LK formula (equation \ref{eq1}), including up to the third harmonics (red curve). Nevertheless, such a non-monotonic change in the QO amplitude is unexpected in a quasi-two-dimensional structure. The effective mass fitting for the $\alpha_1$ frequency [inset of Fig. \ref{fig:4}(d)] reveals a mass smaller than that of the $\alpha_3$ frequency, consistent with the strong QO amplitude observed at these orientations. 

We also measured angular-dependent SdH oscillations while rotating the field in the ab plane in steps of 2.5$^\circ$, as shown in Fig. \ref{fig:5}. Strong SdH oscillations are observed up to $\phi = 70^\circ$. Similarly to the AQO for H $\parallel$ bc (Fig. \ref{fig:4}), the amplitude of the QOs when H $\parallel$ ab does not change monotonically as the field rotates from {\it H} $\parallel b$ to {\it H} $\parallel a$, and the trend is more complex than the AQO in Fig. \ref{fig:4}. In addition, as demonstrated in the contour plot of the FFT amplitude with respect to $\phi$ and $F$ [Fig. \ref{fig:5}(b)], QOs at different field orientations are dominated by different frequencies, corresponding to different Fermi surface cross-sections. Based on the evolution of the QO amplitude and frequencies with increasing $\phi$, the waterfall plot in Fig. \ref{fig:5}(a) can be divided into three regions: (A$'$) for $\phi \leq 7.5^\circ$, the QO amplitude decreases, and the QO signal consists of oscillations of different frequencies ranging from $160\ \text{T}$ to $2.5\ \text{kT}$; (B$'$) for $10^\circ \leq \phi \leq 25^\circ$, the QO amplitude is roughly a constant, and the QOs are dominated by intermediate frequencies ($600\ \text{T} < F < 1500\ \text{T}$); (C$'$) for $\phi \geq 27.5^\circ$, the QO amplitude decreases up to $\phi = 70^\circ$ and the QOs are dominated by small frequencies ($F < 300\ \text{T}$). At $\phi=50^\circ$, the QO amplitude has a singularity (maximum), not easily correlated with other measurements at this angle. The boundaries of the three regions are indicated by dotted lines in Fig. \ref{fig:5}(a). Three specific $\phi$ values are highlighted for later discussion, where the QO amplitude reaches a local minimum or maximum: $\phi_1$ = 7.5$^\circ$ (minimum), $\phi_2$ = 27.5$^\circ$ (maximum), $\phi_3$ = 50$^\circ$ (maximum). 

As with the AQO results in Fig. \ref{fig:4}, the waveform of the SdH oscillations for $27.5^\circ \leq \phi \leq 32.5^\circ$ also appears to deviate from a sinusoidal function [see the magenta box in Fig. \ref{fig:5}(a)], while the periodicity in $1/H$ is preserved. Investigating the SdH oscillations at different temperatures for $\phi_2 = 27.5^\circ$ [Fig. \ref{fig:5}(c)] reveals that they also consist of contributions from higher harmonics [Fig. \ref{fig:5}(d)]. As shown in Fig. \ref{fig:5}(e), the SdH oscillations at 1.8 K (purple symbols) are well-fitted using the LK formula (equation \ref{eq1}), including oscillations up to the third harmonic (red line). The effective mass fitting for the $\alpha_1$ pocket [inset of Fig. \ref{fig:5}(d)] also reveals a smaller mass than that of $\alpha_3$, consistent with the strong QO amplitude observed at these orientations.

Although the waveform in both measurement geometries is explained by the superposition of fundamental oscillations and their higher harmonics, the non-monotonic change in the QO amplitude remains unresolved. A correspondence between the QO frequencies from AQO and the Fermi surface cross-sections needs to be established to address this better. 

We compared the experimental QO frequencies with the cross-sectional areas from DFT calculations, the latter shown as symbols in Fig. \ref{fig:4}(b) and Fig. \ref{fig:5}(b). Good agreement can be achieved between the experiments and the calculations. The QOs with frequencies $\sim$ 160 T and $\sim$ 2100 T are assigned to the cross-sections $\gamma_1$ and $\beta_2$, respectively. Despite a small frequency change, QOs with frequencies of $\sim$ 250 T are related to different Fermi surface cross sections. For {\it H} $\parallel$ bc, these oscillations are assigned to $\alpha_3$ for $\theta \leq 30^\circ$, and to $\alpha_1$ for $\theta > 30^\circ$. For {\it H} in the ac plane, these oscillations are assigned to $\alpha_3$ for $\phi \leq 17.5^\circ$, $\alpha_1$ for $20^\circ\leq \phi \leq 32.5^\circ$, $\beta_1$ for $35^\circ\leq\phi \leq 52.5^\circ$ and $\alpha_2$ for $\phi \geq 47.5^\circ$. In addition, the intermediate oscillation frequencies (600 T $<$ F $<$ 1500 T) for $5^\circ\leq \phi \leq 25^\circ$ are attributed to different cross sections of the $\gamma$ Fermi pocket. When the cross sections that contribute to QOs change due to differences in the underlying band and quasiparticle characteristics (band curvature, quasiparticle effective mass, quantum mobility, etc.), the QO amplitude could undergo an abrupt change, such as the non-monotonic change observed in the QO amplitude of $\mathrm{YbSb_2}$. 

In addition to the QOs from individual Fermi pockets, our analysis also reveals QOs from the magnetic breakdown orbit: $\delta$, $\delta-\alpha_3$, $\delta+\alpha_3$, and $2\delta$. This is consistent with the effective mass analysis shown in the inset of Fig. \ref{fig:2}(c), where $\mu_\delta=0.56\approx(\mu_{\alpha_3}+\mu_{\beta_2})/2=0.58$. $\delta-\alpha_3$ can be seen as the result of the magnetic interaction effect between $\delta$ and $\alpha_3$ \cite{Shoenberg1984}, $\delta+\alpha_3$ represents the magnetic breakdown orbit between $\beta_2$ and $2\alpha_3$, and $2\delta$ is the second harmonic signal of $\delta$.

\begin{figure}
\includegraphics[width=0.48\textwidth]{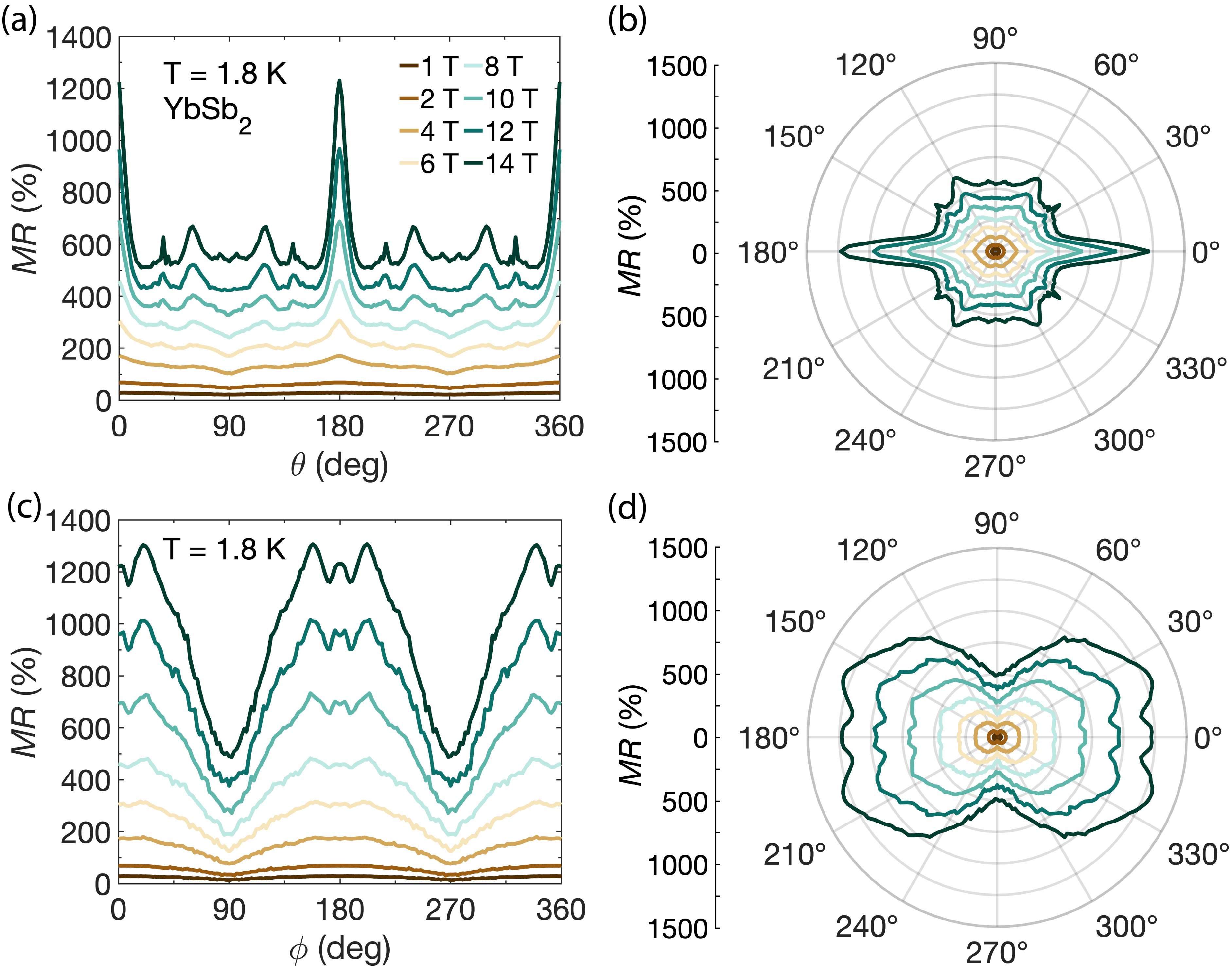} 
\caption{\label{fig:6} Anisotropic magnetoresistance (AMR) of $\mathrm{YbSb_2}$. (a,c) Standard plot of AMR at 1.8 K illustrating the minimum/maximum of the AMR for the magnetic field in the bc plane (a, $\theta$) and the ab plane (c,  $\phi$). (b,d) Polar plot of the AMR at 1.8 K for the magnetic field in the ab plane (b, $\theta$) and the bc plane (d, $\phi$).}
\end{figure}

AMR is a complementary tool for studying the Fermi surface through magnetotransport measurements, with the Fermi surface geometry encoded in the AMR results. We performed AMR measurements on $\mathrm{YbSb_2}$ under different magnetic field orientations at 1.8 K. The measurement geometry for the AMR is the same as that for the AQO measurements [inset of Fig. \ref{fig:4}(b) and Fig. \ref{fig:5}(b)]. Similar to AQO, the AMR does not change monotonically with increasing $\theta$ or $\phi$, and it can also be divided into three regions similar to those in AQO, for both measurement geometries. Fig. \ref{fig:6}(a,b) shows the AMR as a function of $\theta$ (with the magnetic field in the \textit{bc} plane) at 1.8 K. When $\theta$ increases, the AMR (A) decreases significantly for $\theta \leq 20^\circ$, where the AQO amplitude also decreases; (B) reaches a local maximum at $\theta = 36^\circ$ when $20^\circ < \theta \leq 36^\circ$, which is consistent with the maximum amplitude at $\theta_2~=~35^\circ$ in region B; and (C) the AMR decreases slightly for $\theta > 36^\circ$, with a hump at $\theta_3~=~60^\circ$. This angle is higher than the maximum value $\theta~=~45^\circ$ where AQO are still observed. Figure \ref{fig:6}(c,d) shows the AMR as a function of $\phi$ (with magnetic field in the ab plane) at 1.8 K. Three regions can be identified where AMR  (A$'$) decrease for $\phi \leq 8^\circ$, while the QO amplitude also decreases; (B$'$) increases for $8^\circ < \phi \leq 22^\circ$ to a maximum at $\phi = 22^\circ$, which is close to the boundary between regions B$'$ and C$'$ ($\phi_2 \sim 27.5^\circ$) in AQO; and (C$'$) decreases for $\phi \geq 24^\circ$, with a hump at $\phi=48^\circ$, which matches the decrease in the AQO amplitude of region C$'$ and the maximum QO amplitude at $\phi_3 = 50^\circ$. The consistency between the angular dependence of AQO and AMR confirms that the Fermi surface morphology is the underlying reason for the non-monotonic changes in the AQO amplitude.





\section{Discussion}

The non-monotonic changes in the angle-dependent quantum magnetotransport phenomena (AQO) and classical magnetotransport phenomena (AMR) highlight the role of the underlying complex Fermi surface in the magnetotransport properties of $\mathrm{YbSb_2}$. Due to the multi-band nature of $\mathrm{YbSb_2}$, a quantitative analysis of the AMR is very challenging. Instead, we can qualitatively interpret the AMR alongside the AQO. Although quantum mobilities in quantum oscillations and classical mobilities in magnetoresistance refer to two different scattering processes, they both reflect the properties of the underlying Fermi surface and should qualitatively follow the same trend with respect to magnetic field orientations. When the quantum/classical mobility decreases/increases, the amplitude of the AQO/AMR decreases/increases, respectively. This is indeed reflected in the qualitative behavior observed in AQO and AMR, as discussed in the previous section.

The quantum oscillation frequencies from our QO results are well matched by the DFT values, where we identified QOs from the Dirac nodal line pockets ($\alpha_1, \alpha_2$, and $\alpha_3$) and the magnetic breakdown orbits ($\delta$) in the complex FFT frequency spectrum of YbSb$_2$. This indicates that the underlying quasiparticles from the Dirac nodal lines contribute to electrical transport in the normal state of $\mathrm{YbSb_2}$. The electrical transport from these quasiparticles could explain the large, non-saturating MR in the normal state, as shown in Fig. \ref{fig:1_2}(b). 
The electronic band structure of YbSb$_2$ also consists of trivial electronic bands, and the QOs from these bands are observed in the SdH oscillations. The exact bands that are relevant in the formation of Cooper pairs in the superconducting state are currently unknown. By assuming that the quasiparticles from the Dirac nodal lines contribute to superconductivity, we propose a scenario where these quasiparticles are related to the unusual type I superconductivity in YbSb$_2$. The coherence length $\xi$ is reported to be much larger than the penetration depth $\lambda$ in $\mathrm{YbSb_2}$ \cite{Zhao2012}, consistent with type I superconductivity. The coherence length in a superconductor depends on the coherence length for a pure material $\xi_0$ and the mean free path of the carriers $l$: 

\begin{equation}
\frac{1}{\xi}=\frac{1}{\xi_0}+\frac{1}{l}
\end{equation}

Compared to quasiparticles from trivial bands, the relativistic quasiparticles from topological bands have smaller effective masses and higher mobilities \cite{Liang2015,Hu2019}, which corresponds to a longer mean free path {\it l}. The coherence length of the Cooper pairs formed by these relativistic particles will be longer than those formed by non-relativistic particles, thus favoring type I superconductivity. The detailed superconducting pairing mechanism, its relationship to band topology, and their implications for topological superconductivity in $\mathrm{YbSb_2}$ deserve extensive theoretical and experimental studies in the future, and shall not be the focus of the current study.


\section{Conclusions}
In summary, we report large anisotropic magnetoresistance and strong QOs in the quasi-two-dimensional Dirac nodal line superconductor $\mathrm{YbSb_2}$. The complex Fermi surface geometry and the change in the underlying Fermi surface cross sections result in non-monotonic angular dependence of the AMR and AQOs. By matching the experimental QO frequencies to the calculated extremal cross-sectional areas, we identified QOs from the Dirac nodal line and the magnetic breakdown orbit associated with the Dirac nodal line. Our results suggest that the non-trivial band topology is linked to type I superconductivity in $\mathrm{YbSb_2}$, making it a good material platform to explore the possible interplay between superconductivity and band topology.

\begin{table*}

    \centering
    \begin{ruledtabular}
    \begin{tabular}{cccccccc}
    &\multicolumn{7}{c}{$\mathrm{YbSb_2}$ H $\parallel$b}\\
     Index&$F(T)$&$T_D(K)$&$m_{\texttt{eff}}$/$m_0$&$\tau_q(ps)$&$\mu_q(cm^2V^{-1}s^{-1})$   &$k_F(\AA)$&$v_F(10^6ms^{-1})$  \\ \hline
     $\gamma_2$&163.1&4.6&0.31&0.266&1472&0.070&0.256   \\
     $\alpha_3$&236.2&6.8&0.26&0.180&1110&0.085&0.345  \\
     $\delta$&1419&4.7&0.60&0.259&769&0.208&0.406 \\
     $\beta_2$&2141&3.53&0.89&0.344&677&0.255&0.330 \\
     $\alpha_3$ dHvA&236.2&3.4&0.15&0.355&4120&0.085&0.647  \\
     \end{tabular}
     
    \caption{\label{table:1}Parameters extracted from QOs of $\mathrm{YbSb_2}$ for $\textbf{H}$ $\parallel$ $\textbf{b}$. $F$ is the oscillation frequency; T$_D$ is the Dingle temperature;  $\tau_q$ is the quantum lifetime; $\mu_q$ is the quantum mobility; $k_F$ and $v_F$ are Fermi vector and Fermi velocity; $\phi_B$ is the Berry phase; g is the Landé-g factor.}
    \end{ruledtabular}
     
    \label{tab:my_label}
    
\end{table*}

\begin{acknowledgments}
This work was primarily supported by the Department of Defense Air Force Office of Scientific Research under Grant No. FA9550-21-1-0343. RAM, SM and EM acknowledge partial support from the Kavli Foundation 
as part of the Super C collaboration. The work at Los Alamos National Laboratory was carried out under the auspices of the US Department of Energy (DOE) National Nuclear Security Administration under Contract No. 89233218CNA000001, and was supported by the Los Alamos National Laboratory (LANL) LDRD Program, and in part by the Center for Integrated Nanotechnologies, an Office of Science User Facility operated by the U.S. Department of Energy (DOE) Office of Science, in partnership with the LANL Institutional Computing Program for computational resources. The ARPES work described in this paper was performed at the Canadian Light Source, a national research facility of the University of Saskatchewan, which is supported by the Canada Foundation for Innovation (CFI), the Natural Sciences and Engineering Research Council (NSERC), the Canadian Institutes of Health Research (CIHR), the Government of Saskatchewan, and the University of Saskatchewan. 

\end{acknowledgments}


  \pagebreak

\newpage

\begin{appendices}

\setcounter{figure}{0}
\section{4$f$ electronic bands in ARPES measurements}

\begin{figure}
\includegraphics[width=0.48\textwidth]{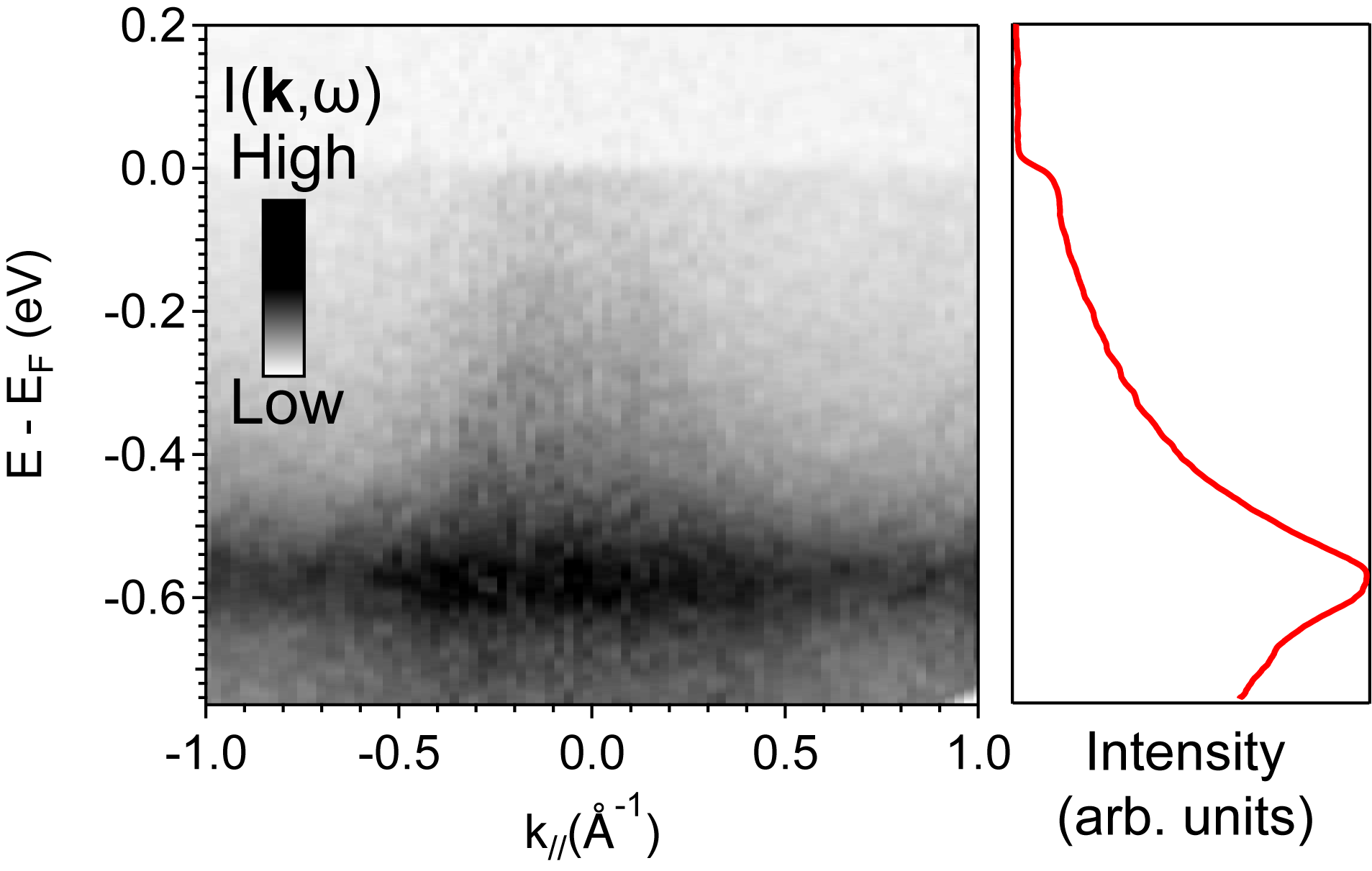}
\renewcommand{\thefigure}{S\arabic{figure}}
\caption{\label{fig:SI1} 
Electronic band structure of $\mathrm{YbSb_2}$ obtained from ARPES measurements. The left panel is the band structure obtained with photons of 87 eV and $p$-polarization at 15 K. The right panel is the corresponding energy distribution curve integrated across the presented momentum slice. 
}
\end{figure}

ARPES measurements on YbSb$_2$ shown in Fig.~\ref{fig:SI1} clearly reveal the non-dispersive 4$f$ bands at around $E-E_{\rm F}$ = -0.6~eV, contributing to the prominent peak in the integrated energy distribution curve. This feature is used to gauge the orbital energy-dependent correction in DFT calculations. Further, dispersive bands along the in-plane momentum direction crossing the Fermi level are observed to form a hole-like pocket, consistent with the DFT+$U$ prediction around $\Gamma$.

\end{appendices}
\end{document}